\documentclass[aps, prd, reprint, twocolumn, superscriptaddress]{revtex4-1}
\usepackage{amsmath}
\usepackage{amssymb}
\usepackage{graphicx}
\usepackage{color}
\usepackage{times}
\usepackage{euscript}
\usepackage{amsthm}
\usepackage{amsfonts}
\usepackage[english]{babel}
\usepackage{epstopdf}
\usepackage{multirow,makecell,array}
\usepackage{mathtools}
\usepackage{float}
\usepackage[dvipsnames]{xcolor}
\usepackage{ulem}

\begin{document}

\title{Schwinger particle production: Rapid switch off of the external field versus dynamical assistance}

\author{I.~A.~Aleksandrov}
\affiliation{Department of Physics, Saint Petersburg State University, Universitetskaya Naberezhnaya 7/9, Saint Petersburg 199034, Russia}
\affiliation{Ioffe Institute, Politekhnicheskaya Street 26, Saint Petersburg 194021, Russia}
\author{D.~G.~Sevostyanov}
\affiliation{Department of Physics, Saint Petersburg State University, Universitetskaya Naberezhnaya 7/9, Saint Petersburg 199034, Russia}
\author{V.~M.~Shabaev}
\affiliation{Department of Physics, Saint Petersburg State University, Universitetskaya Naberezhnaya 7/9, Saint Petersburg 199034, Russia}

\begin{abstract}
We consider the process of electron-positron pair production in the presence of strong electric backgrounds being rapidly switched on and off and examine the total particle yield. For sufficiently sharp field profiles, the particle number can be substantially enhanced. It is demonstrated that this enhancement is quite similar to the phenomenon of dynamical assistance by a weak high-frequency field superimposed on a strong background. Both these mechanisms are analyzed by means of exact numerical computations, worldline instanton approach, and the locally-constant field approximation. We identify the timescale of the switching profile leading to the pair-production enhancement and argue that the particle yield is highly unlikely to be increased by shaping the switch on and off of realistic laser pulses. On the other hand, we confirm that it is feasible to observe the dynamically assisted Schwinger effect by adding a rapidly oscillating field to a strong electric background.
\end{abstract}

\maketitle

\section{Introduction}\label{sec:intro}

One of the most striking phenomena predicted by quantum electrodynamics (QED) is the process of electron-positron pair production from vacuum in intense electromagnetic fields~\cite{sauter_1931, heisenberg_euler, schwinger_1951}. In the strong-field regime, where the external background cannot be treated within perturbation theory (PT), this effect (Schwinger mechanism) still awaits experimental observations. By continually increasing the pulse intensity that can be achieved at the modern laser facilities, experimentalists have stimulated great interest in strong-field QED over the past decades (see, e.g, recent reviews~\cite{xie_review_2017, gonoskov_2022,fedotov_review}), so it is also extremely important for theoretical groups to explore the possibilities for enhancing the particle yield and propose most favorable practical scenarios. Here we will focus on two general ideas.

First, it is known that evaluating the adiabatic number of particles at intermediate times, when the external field is on, one may obtain values that are orders of magnitude larger than the residual (asymptotic) pair yield (see, e.g., Refs.~\cite{fedotov_prd_2011, blaschke_prd_2013, blinne_prd_2014, zahn_2015, otto_jpp_2016}). Although it is the latter which is to be observed in the experiment, one can attempt to maximize it by choosing a more advantageous field profile. Recently, it was suggested that by properly switching the external field off, one can indeed make the large adiabatic particle number observable~\cite{ilderton_prd_2022}. The switching profile should be sufficiently sharp, and its location should be accurately adjusted. This issue was addressed in Ref.~\cite{ilderton_prd_2022} based on the analysis of the electron (positron) number density evaluated nonperturbatively and the total particle yield computed within PT. In the present study, we examine the total number of pairs without perturbative expansions.

Second, in Ref.~\cite{schuetzhold_prl_2008} it was demonstrated that the number of pairs created from vacuum can be drastically enhanced by superimposing a weak rapidly oscillating pulse on a strong quasistatic external background. Although the former has a small amplitude, it provides a high-frequency harmonic which efficiently stimulates the pair production process. As this so-called dynamically assisted Schwinger effect (DASE) appears to be very promising from the experimental viewpoint, it has been explored in numerous investigations (see, e.g., Refs.~\cite{orthaber_2011, fey, kohlfuerst_prd_2013, linder_prd_2015, akal_prd_2014, hebenstreit_plb_2014, otto_plb_2015, panferov_epjd_2016, aleksandrov_prd_2018, otto_epja_2018, sitiwaldi_plb_2018, torgrimsson_prd_2018, huang_prd_2019, chavez_prd_2019, olugh_plb_2020, taya_prr_2020, li_prd_2021}).

Here we discuss how the two aforementioned phenomena enhancing the particle yield are related. The adiabatic number of $e^+e^-$ pairs evaluated at intermediate times is the number of physical pairs which would be measured if the external field were suddenly turned off at that time~\cite{ilderton_prd_2022}. This suggests that it is the high-frequency mode induced by the rapid switching profile which multiplies the particle yield. As will be demonstrated in this paper, this mechanism is closely related to the DASE. We will scrutinize this relationship and identify the common patterns of the two effects. In order to provide a connection with the experimental setups, we will find out which values of the field parameters lead to the enhancement of the particle number and assess the feasibility of the corresponding practical scenarios.

To calculate the number of $e^+e^-$ pairs, we will employ our numerical procedure using the quantum kinetic equations (QKE)~\cite{aleksandrov_qke_2020, sevostyanov_prd_2021} and also verify the results by our computational approach based on the Furry-picture quantization in momentum space~\cite{aleksandrov_prd_2016, aleksandrov_prd_2017_2, aleksandrov_kohlfuerst}. These nonperturbative approaches neglect the radiative corrections but fully incorporate the interaction between the quantized Dirac field and the classical background. For simplicity, we assume the external field to depend solely on time (the corresponding temporal profile is treated without any additional approximations). Besides, we utilize the worldline instanton technique~\cite{dunne_prd_2005, dunne_prd_2006} and the so-called locally-constant field approximation (LCFA)~\cite{bunkin_tugov, narozhny_pla_2004, dunne_prd_2006, hebenstreit_prdr_2008, bulanov_prl_2010, gavrilov_prd_2017, sevostyanov_prd_2021, aleksandrov_prd_2019_1, aleksandrov_symmetry_2022}. The validity of the latter approach to calculating the total particle yield was recently examined in Ref.~\cite{sevostyanov_prd_2021} (see also Refs.~\cite{aleksandrov_prd_2019_1, aleksandrov_symmetry_2022}). Finally, to further illustrate the main patterns of the enhancement mechanisms, we will perform the analysis of the Fourier transforms of the external electric backgrounds.

The paper is organized as follows. In Sec.~\ref{sec:setup} we describe the field configurations to be investigated. Our exact numerical procedure based on the QKE is outlined in Sec.~\ref{sec:qke}. In Sec.~\ref{sec:lcfa} we present the general expressions obtained within the LCFA. Section~\ref{sec:results} contains our analysis of the enhancement mechanisms. The experimental prospects regarding particle enhancement are discussed in Sec.~\ref{sec:experiment}. We conclude in Sec.~\ref{sec:conclusions}.

Throughout the text, we employ the units where $\hbar = c = 1$ and the electric charge $e<0$ obeys $\alpha = e^2/(4\pi)$ ($\alpha \approx 1/137$ is the fine-structure constant). 

\begin{figure*}[t]
  \center{\includegraphics[height=0.35\linewidth]{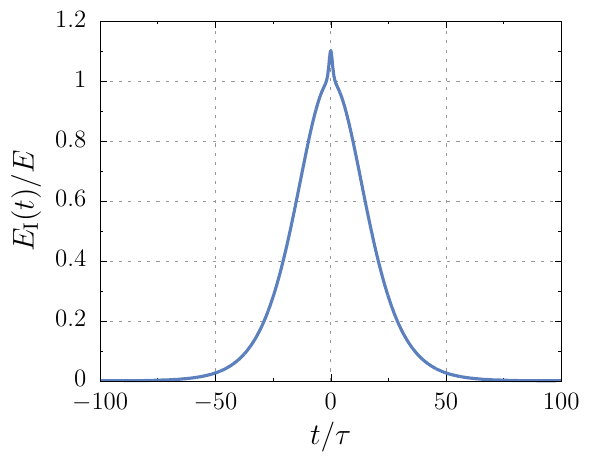}~~~~~~~~~\includegraphics[height=0.35\linewidth]{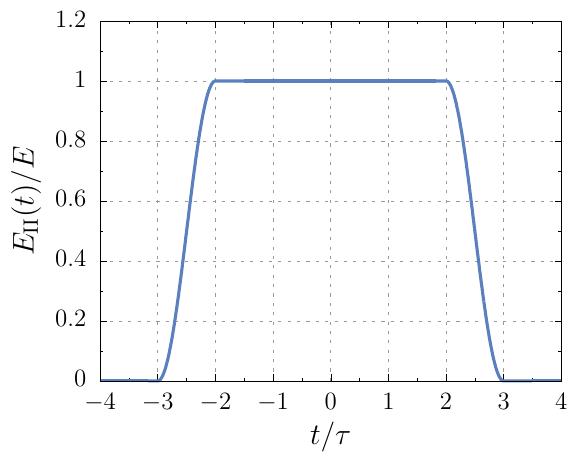}}
  \caption{Field  profiles~\eqref{eq:E1} and \eqref{eq:E2} for specific field parameters: $\varepsilon = 0.1 E$, $T = 10 \tau$ [$E_\text{I} (t)$, left] and $T=10 \tau$ [$E_\text{II} (t)$, right].}
  \label{fig:profiles}
\end{figure*}

\section{Setup. External electric field} \label{sec:setup}

We choose the classical external background (in $3+1$ dimensions) in the form of a linearly polarized electric field $E(t)$ vanishing at $t \to \pm \infty$. The main patterns of the DASE will be discussed in the case of a combination of two Sauter pulses:
\begin{equation}
E_\text{I} (t) = \frac{E}{\cosh^2 (t/T)} + \frac{\varepsilon}{\cosh^2 (t/\tau)}.
\label{eq:E1}
\end{equation}
The strong pulse parameters obey $|e|ET \gtrsim m$, so it cannot be treated within PT. Varying the weak pulse duration $\tau$, we will identify the onset of dynamical assistance and thoroughly examine the $\tau$ dependence of the total number of pairs.

The role of the switching profile and possible enhancement due to the switching effects will be analyzed for the following external field configuration:
\begin{widetext}
\begin{equation}
E_\text{II} (t) = 
\begin{cases}
E & \text{if}~~|t| \leq T/2, \\
E \cos^2 \big [ \frac{\pi}{2\tau} (|t| - T/2) \big ] &\text{if}~~T/2 < |t| \leq T/2 + \tau, \\
0 &\text{otherwise}.
\end{cases} \label{eq:E2}
\end{equation}
\end{widetext}
For simplicity it is chosen in a symmetric form. Here $T$ is the temporal extent of the plateau region, where the external field is constant, and $\tau$ governs the duration of the smooth switch-on and switch-off parts. The field profiles~\eqref{eq:E1} and \eqref{eq:E2} are displayed in Fig.~\ref{fig:profiles}.

Since the external field is assumed to be spatially homogeneous, the total number of pairs will be proportional to the volume of the system, $N \sim V$, so in what follows we will present the results in terms of the pair yield per unit volume $\nu = N/V$. The corresponding numbers will also carry the subscripts ``I'' and ``II''.

\section{Quantum kinetic equations} \label{sec:qke}

The pair production process in the case of a spatially homogeneous external field can be nonperturbatively described by means of the so-called quantum kinetic equations (QKE) derived first in Refs.~\cite{GMM, schmidt_1998, kluger_1998} and generalized in Refs.~\cite{pervushin_2006, filatov_2006} (see also Refs.~\cite{aleksandrov_qke_2020,aleksandrov_kudlis_klochai_2024} and references therein).

The external electric field $E(t)$ corresponds to the vector potential $A(t)$ [$E(t) = -\partial A/\partial t$] and is assumed to vanish outside the interval $t \in [t_\text{in}, \, t_\text{out}]$ (in the case of Sauter pulses $t_\text{in/out} \to \mp \infty$). Let us introduce the electron number density in momentum space:
\begin{equation}
  f(\boldsymbol{p}) = \frac{(2\pi)^3}{V} \frac{dN_{\boldsymbol{p}}}{d\boldsymbol{p}}.
  \label{eq:f_dist_def}
\end{equation}
The total number of pairs per unit volume is then given by
\begin{equation}
\nu = \frac{2}{(2\pi)^3} \int \! f(\boldsymbol{p}) \, d\boldsymbol{p}.
  \label{eq:f_nu}
\end{equation}
The factor of 2 takes into account the spin degeneracy: the spin-resolved quantity~\eqref{eq:f_dist_def}, in fact, does not depend on the spin state in the case of a linearly polarized external electric field. The integration in Eq.~\eqref{eq:f_nu} is practically two-dimensional due to the azimuthal symmetry of the setup. In order to obtain the electron spectrum $f(\boldsymbol{p})$, one can (numerically) solve the following system of differential equations:
\begin{eqnarray}\label{eq:cauchy_1}
\dot{f}(\boldsymbol{p},t) &=& \frac{1}{2}\lambda (\boldsymbol{p},t) u(\boldsymbol{p},t),\\
\dot{u}(\boldsymbol{p},t) &=& \lambda (\boldsymbol{p},t) \bigl[1-2f(\boldsymbol{p},t)\bigr] - 2\omega (\boldsymbol{p},t) v(\boldsymbol{p},t), \label{eq:cauchy_2} \\
\dot{v}(\boldsymbol{p},t) &=& 2\omega (\boldsymbol{p},t) u(\boldsymbol{p},t). \label{eq:cauchy_3}
\end{eqnarray}
Here the initial conditions read $f(\boldsymbol{p}, t_{\text{in}}) = u(\boldsymbol{p}, t_{\text{in}}) = v(\boldsymbol{p}, t_{\text{in}}) = 0$ for all $\boldsymbol{p}$. The final values of $f(\boldsymbol{p},t)$ yield the momentum distribution, $f(\boldsymbol{p},t_\text{out}) = f(\boldsymbol{p})$. The coefficients in Eqs.~\eqref{eq:cauchy_1}--\eqref{eq:cauchy_3} are defined via
\begin{eqnarray}
  && \lambda (\boldsymbol{p},t) = \frac{eE(t)\mu (\boldsymbol{p})}{\omega^2(\boldsymbol{p}, t)}, \\
  && \mu (\boldsymbol{p}) = \sqrt{m^2+\boldsymbol{p}_{\perp}^2}, \\
  && \omega(\boldsymbol{p}, t) = \sqrt{\mu^2 (\boldsymbol{p}) + [p_{\parallel}-eA(t)]^2},
\end{eqnarray}
where $p_{\parallel}$ and $\boldsymbol{p}_{\perp}$ are the longitudinal (with respect to the electric field) and transverse momentum components, respectively.

The QKE are solved numerically for given $p_\parallel$ and $|\boldsymbol{p}_\perp|$. After that, we integrate the spectrum according to Eq.~\eqref{eq:f_nu} ($d\boldsymbol{p} \to 2 \pi |\boldsymbol{p}_\perp| d |\boldsymbol{p}_\perp| dp_\parallel$).

\section{Locally-constant field approximation} \label{sec:lcfa}

The idea of the LCFA is to employ a closed-form expression for the total number of pairs produced per unit volume and time in the presence of a constant electromagnetic field. This reads~\cite{nikishov_constant}
\begin{equation}
  \frac{dN}{dt d\boldsymbol{x}} \, [\mathcal{E}, \mathcal{H} ] = \frac{e^2 \mathcal{E} \mathcal{H}}{4\pi^2} \coth{\frac{\pi \mathcal{H}}{\mathcal{E}}} \, \mathrm{exp} \bigg ( \! - \frac{\pi m^2}{|e|\mathcal{E}} \bigg ),
  \label{eq:LCFA_gen}
\end{equation}
where
\begin{eqnarray}
  \mathcal{E} &=& \sqrt{\sqrt{\mathcal{F}^2 + \mathcal{G}^2} + \mathcal{F}},\label{eq:E_cal} \\
  \mathcal{H} &=& \sqrt{\sqrt{\mathcal{F}^2 + \mathcal{G}^2} - \mathcal{F}}.\label{eq:H_cal}
\end{eqnarray}
Here the invariant quantities $\mathcal{F}$ and $\mathcal{G}$ are defined in terms of the electric and magnetic field vectors via $\mathcal{F} = (\boldsymbol{E}^2 - \boldsymbol{H}^2)/2$ and $\mathcal{G} = \boldsymbol{E}\cdot \boldsymbol{H}$. Accordingly, $\mathcal{E}$ and $\mathcal{H}$ are the magnitudes of the electric and magnetic field components, respectively, in the frame where they are parallel to each other. The Schwinger critical value of the electric field strength is $E_\text{c} = m^2/|e|$.

Within the LCFA, one plugs the actual spatiotemporal dependence of the external inhomogeneous field into Eq.~\eqref{eq:LCFA_gen} and integrates it then over $t$ and $\boldsymbol{x}$ (see Refs.~\cite{bunkin_tugov, narozhny_pla_2004, dunne_prd_2006, hebenstreit_prdr_2008, bulanov_prl_2010, gavrilov_prd_2017, sevostyanov_prd_2021, aleksandrov_prd_2019_1, aleksandrov_symmetry_2022}). In the case of a spatially uniform electric field $E(t)$, this procedure yields the following expression for the total number of pairs per unit volume:
\begin{equation}
 \nu^{\text{(LCFA)}} = \frac{e^2}{4\pi^3} \int \! dt \,  E^2(t) \, \mathrm{exp} 
  \left( \! - \frac{\pi m^2}{|eE(t)|} \right).
  \label{eq:LCFA_uniform}
\end{equation}

In what follows, we will perform the calculations using the specific forms~\eqref{eq:E1} and \eqref{eq:E2}. Although in the case of two Sauter pulses~\eqref{eq:E1}, the LCFA prediction~\eqref{eq:LCFA_uniform} is nonlinear with respect to the field parameters, employing the function $E_\text{II} (t)$ leads to the following result:
\begin{equation}
\nu^{\text{(LCFA)}}_{\text{II}} = [T + f(E/E_\text{c}) \tau ] g(E), \label{eq:LCFA_nu2}
\end{equation}
where
\begin{equation}
g(E) = \frac{e^2 E^2}{4\pi^3} \mathrm{exp} 
  \left( \! - \frac{\pi m^2}{|eE|} \right),
  \label{eq:LCFA_g}
\end{equation}
which represents the number of pairs created in a constant electric field $E$ per unit volume and time. We also introduced
\begin{equation}
f (\xi) = \frac{4}{\pi} \int \limits_0^{\pi/2} dx \cos^{4} x \ \mathrm{exp}   \left( \! - \frac{\pi}{\xi} \tan^2 x \right). \label{eq:LCFA_f2}
\end{equation}
Accordingly, for given field strength $E$, the particle number evaluated within the LCFA linearly depends on the parameters $T$ and $\tau$. We also note that $f(\xi) \approx (2/\pi)\sqrt{\xi}$ for $\xi \ll 1$ (see Appendix~\ref{sec:app_f}).

Equation~\eqref{eq:LCFA_nu2} contains $g(E)$, which reflects the fact that a constant electric background generates $e^+e^-$ pairs in the tunneling (nonperturbative) regime. The LCFA is valid only in the domain of sufficiently large field amplitudes and pulse durations. The approximate predictions and the results of exact numerical computations will be discussed next.

\section{Results. Enhancement mechanisms} \label{sec:results}

Here we will first evaluate the total number of pairs produced for the two scenarios described above as a function of~$\tau$, which represents there the weak-pulse duration and the duration of the switch-on and switch-off parts, respectively. To gain additional insights into the enhancement mechanisms, we will also analyze the Fourier spectra of the corresponding electric backgrounds~\eqref{eq:E1} and \eqref{eq:E2}.

\subsection{Total particle yield. Dynamical assistance} \label{sec:yield_dase}

\begin{figure}[t]
  \center{\includegraphics[width=0.999\linewidth]{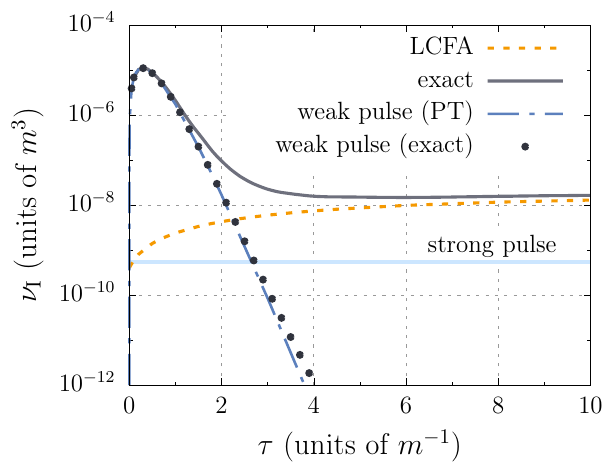}}
  \caption{Total number of $e^+e^-$ pairs per unit volume created by the external field~\eqref{eq:E1} as a function of the weak-pulse duration $\tau$ ($E = 0.2 E_\text{c}$, $T = 20 m^{-1}$, $\varepsilon = 0.05 E_\text{c}$). The results are obtained within the LCFA and by means of our exact numerical technique based on the QKE. The dash-dotted line and thick dots correspond to the particle yield produced by the weak pulse only [second term in Eq.~\eqref{eq:E1}].}
  \label{fig:dase}
\end{figure}

In Fig.~\ref{fig:dase} we present the total number of pairs as a function of the weak-pulse duration $\tau$ in the case of the external field~\eqref{eq:E1}. This example was obtained for $E = 0.2 E_\text{c}$, $T = 20 m^{-1}$, and $\varepsilon = 0.05 E_\text{c}$. Before we discuss the dynamical assistance, which occurs for sufficiently small $\tau$, let us briefly outline the main patterns concerning the LCFA, which clearly does not capture the DASE. For sufficiently large $\tau$, the exact values of $\nu_\text{I}$ (gray solid line) almost coincide with $\nu_\text{I}^{\text{(LCFA)}}$ (dashed line)~\cite{lcfa_comment}. We note that although the LCFA is quite accurate in the case of two pulses, it completely fails to describe the pair production process in the field of the weak pulse {\it alone} (e.g., for $\tau = 10m^{-1}$ the LCFA underestimates the particle yield by approximately 14 orders of magnitude). With decreasing $\tau$, the LCFA prediction decreases, which is evident already from the general LCFA expression~\eqref{eq:LCFA_uniform} as for smaller $\tau$ the local values $E(t)$ decrease. Since the LCFA incorporates only the tunneling pair-production mechanism, this approximation cannot capture the dynamical processes.

Figure~\ref{fig:dase} clearly demonstrates that it is the dynamical effect which enhances the exact particle number. It turns out that the number of pairs in the combination of two Sauter pulses can substantially exceed each of the two individual contributions. This result constitutes the DASE.

Nevertheless, once the weak-pulse duration $\tau$ becomes too small, the individual weak pulse may produce almost the same number of pairs as the two pulses generate together. In this domain, the weak pulse does not assist the strong one but rather governs the pair-production process unaided. In order to simplify the analysis and benchmark the nonperturbative methods, we also employ PT in the case of the individual weak pulse (dash-dotted line in Fig.~\ref{fig:dase}). The total number of pairs per unit volume to the second order in the external-field amplitude $\varepsilon$ reads (see Appendix~\ref{sec:app_PT})
\begin{equation}
\nu_\text{weak}^{\text{(PT)}} = \frac{e^2 \varepsilon^2 \tau^4}{4\pi} \int d\boldsymbol{p} \, \frac{m^2 + \boldsymbol{p}_\perp^2}{p_0^2} \, \frac{1}{\sinh^2 (\pi \tau p_0)},
\label{eq:weak_PT}
\end{equation}
where $\boldsymbol{p}_\perp$ is the transverse momentum component and $p_0 = \sqrt{m^2 + \boldsymbol{p}^2}$. The expression~\eqref{eq:weak_PT} is valid only for large values of the Keldysh parameter $\gamma_\varepsilon \equiv m/|e\varepsilon \tau| \gg 1$. For instance, for $\tau = 5m^{-1}$ the PT approach underestimates the exact yield by a factor of $\sim 20$ and for $\tau = 10m^{-1}$ the PT result is already about ten orders of magnitude smaller.

Finally, for very small $\tau$ ($m\tau \ll 1$), one can expand Eq.~\eqref{eq:weak_PT} in powers of $\tau$ and obtain the following leading-order contribution (see Appendix~\ref{sec:app_PT}):
\begin{equation}
\nu_\text{weak}^{\text{(PT)}} \approx \frac{1}{9\pi} \, \frac{\varepsilon^2}{E_\text{c}^2} \, (m\tau) m^3.
\label{eq:weak_PT_lin}
\end{equation}
Accordingly, in the regime $m\tau \ll 1$ the number of pairs produced by the weak field linearly depends on the pulse duration and vanishes at $\tau = 0$.

\begin{figure}[t]
  \center{\includegraphics[width=0.93\linewidth]{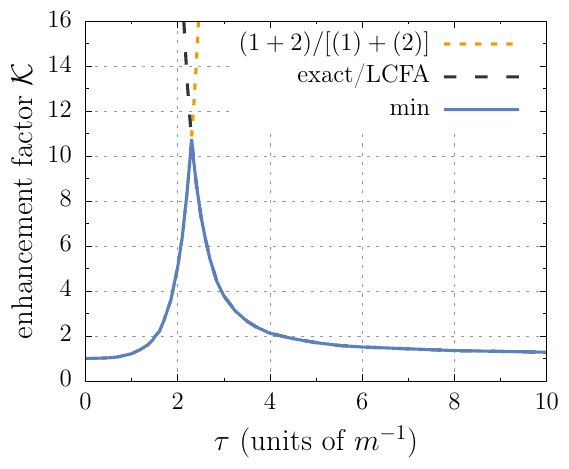}}
  \caption{Enhancement factor defined in Eq.~\eqref{eq:dase_K} as a function of $\tau$ for $E = 0.2 E_\text{c}$, $T = 20 m^{-1}$, and $\varepsilon = 0.05 E_\text{c}$.}
  \label{fig:dase_enhancement}
\end{figure}

In order to quantitatively estimate the efficiency of dynamical assistance, one can first calculate the ratio $\nu_\text{I}/(\nu_\text{strong} + \nu_\text{weak})$, which shows the relative increase in the particle yield arising when the strong and weak pulses are combined. However, this quantity becomes irrelevant in the domain of large $\tau$ since in this case the number of pairs can be evaluated simply by means of the LCFA, i.e., by taking into account the tunneling mechanism [note that in the tunneling regime for $T=\tau$, the ratio is always large due to the highly nonlinear dependence of the particle number on the field amplitude, see Eq.~\eqref{eq:LCFA_uniform}]. Accordingly, we suggest that one relies on the enhancement factor defined via
\begin{equation}
\mathcal{K} = \mathrm{min}~ \Bigg \{ \frac{\nu_\text{I}}{\nu_\text{strong} + \nu_\text{weak}},~\frac{\nu_\text{I}}{\nu^{\text{(LCFA)}}_\text{I}} \Bigg \}.
\label{eq:dase_K}
\end{equation}
In Fig.~\ref{fig:dase_enhancement} we depict the enhancement factor $\mathcal{K}$ as a function of~$\tau$. In our example, the presence of the short weak pulse can enhance the total number of particles by one order of magnitude (in principle, by varying the field parameters, one can achieve a stronger enhancement, see, e.g., Ref.~\cite{aleksandrov_prd_2018}). The most important point for the assessment of the experimental prospects is that the onset of the DASE corresponds to sufficiently small values of $\tau$: for $\varepsilon \ll E$, the combined Keldysh parameter $\gamma_\text{c} \equiv m/|eE\tau|$ should be sufficiently large, $\gamma_\text{c} \gtrsim 1$~\cite{schuetzhold_prl_2008}. The peak in Fig.~\ref{fig:dase_enhancement} is located at $\tau_* \approx 2.3 m^{-1}$. Including the weak-pulse amplitude $\varepsilon$, which in our case is not that small compared to $E$, we obtain $m/|e(E + \varepsilon) \tau_*| \approx 1.7$. Let us underline that the condition $\gamma_\text{c} \gtrsim 1$ reflects the onset of the DASE, but one has to make sure that the nonperturbative nature of the pair production process is preserved. Since the strong (weak) pulse operates in the tunneling (perturbative) regime, the weak pulse should only assist the particle production mechanism. 

Dynamical assistance can be viewed as a process of additional photon absorption galvanizing the tunneling mechanism. For this to happen, the energy spectrum of the weak pulse should be sufficiently broad, so that it contains large $\omega$, i.e., $\tau$ should be sufficiently small (note that the energy spectrum of a Sauter pulse is centered at $\omega = 0$). This will be examined in more detail in Sec.~\ref{sec:fourier}. Instead of two Sauter pulses, one can also analyze a combination of two oscillating pulses with a low and high frequency, respectively (see, e.g., Refs.~\cite{linder_prd_2015, akal_prd_2014, hebenstreit_plb_2014, otto_plb_2015, panferov_epjd_2016, aleksandrov_prd_2018, otto_epja_2018, sitiwaldi_plb_2018, olugh_plb_2020, li_prd_2021}). Although this configuration appears to be more relevant from the experimental point of view, the main patterns of the DASE are the same. For instance, the number of particles as a function of the weak-pulse frequency $\omega$ has qualitatively the same shape as does the $\tau$-dependence revealed in Fig.~\ref{fig:dase} once one maps $\omega \sim 1/\tau$~\cite{kohlfuerst_prd_2013, kohlfuerst_arxiv_2022}. In the case of oscillating pulses, the onset of the DASE is governed by the combined Keldysh parameter $\gamma_\text{c} = m\omega/|eE|$.

\begin{figure}[t]
  \center{\includegraphics[width=0.99\linewidth]{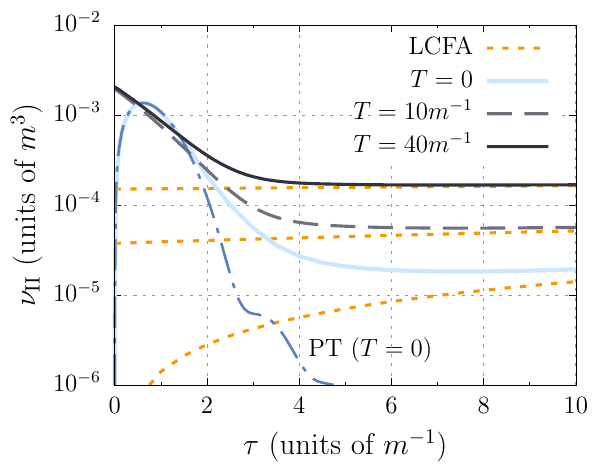}}
  \caption{Total number of $e^+e^-$ pairs in the case of the external field~\eqref{eq:E2} as a function of $\tau$ for various values of $T$ ($E = 0.5 E_\text{c}$). The dashed lines correspond to the LCFA predictions. The total number of pairs for $T=0$ is also evaluated within PT (dash-dotted line).}
  \label{fig:switch_off}
\end{figure}

\subsection{Total particle yield. Switch-off effects} \label{sec:yield_switching}

We now turn to the study of the enhancement mechanism due to a sharp switch on and off of the external field. The main idea is to make use of the fact that the adiabatic particle number at intermediate times can be orders of magnitude larger than the final particle yield. To maximize the number of pairs, one can rapidly turn the field off, so that the large adiabatic number transforms to a physically observable quantity~\cite{ilderton_prd_2022}. This enhancement mechanism can be efficient only if the corresponding smooth switch off is sufficiently rapid. Here our main goal is to analyze the timescale of the switching profile required for the pair-production enhancement and to assess then the possible experimental implementations of this mechanism. It turns out that one can substantially deepen understanding of this scenario by comparing it with the DASE. As will be seen below, the two phenomena are closely related.

We will examine the external-field configuration given in Eq.~\eqref{eq:E2}, which represents a rectangular-like pulse with smooth switch-on and switch-off parts of duration $\tau$ each. First, we note that in the case of a {\it sharp} rectangular profile with parameters $T$ and $\tau = 0$, the adiabatic number of particles at time instant~$t$ ($|t| < T/2$) obviously coincides with the final pair yield for the pulse with duration $T' = t + T/2$. The adiabatic particle number is equally nonphysical as the sudden switch on and off of the rectangular pulse for $\tau = 0$. To make the field configuration more realistic, we consider nonzero values of $\tau$.

In Fig.~\ref{fig:switch_off} we present the total number of pairs produced as a function of $\tau$ for $E = 0.5 E_\text{c}$ and various durations of the plateau region $T$. Here we observe that the $\tau$ dependence is very similar to that displayed in Fig.~\ref{fig:dase}. According to Eq.~\eqref{eq:LCFA_nu2}, the LCFA prediction is linear and decreases with decreasing $\tau$ (the dashed lines have a log-like shape although they may seem straight in the graph). Within the LCFA, for $\tau = 0$ we obtain the constant-field result~\eqref{eq:LCFA_g} multiplied by the rectangular-pulse duration $T$. For large values of $\tau$, the exact particle yield can be accurately obtained within the LCFA since the switch-on and switch-off parts can be treated by means of the LCFA if $|eE|^{3/2} \tau \gg m^2$~\cite{sevostyanov_prd_2021}. We also point out that the quality of the LCFA is better for large $T$ since a broader plateau region yields a larger amount of particles and reduces a {\it relative} contribution of the switch-on and switch-off parts.

In the case of $T=0$, the field profile turns to a $\cos^2$ pulse, whose shape looks similar to that of the Sauter pulse. In this case, we also employ PT and reproduce the exact result in the region $|eE|\tau \ll m$, i.e., the dash-dotted line in Fig.~\ref{fig:switch_off} coincides with the solid light-blue one for small~$\tau$. (We note that the PT approach is much less accurate here than in the context of the DASE because the amplitude $E$ of the rectangular pulse is ten times larger than $\varepsilon$ in Fig.~\ref{fig:dase}.) If the pulse duration $\tau$ becomes extremely small, $m\tau \ll 1$, the PT result exhibits a linear behavior (see Appendix~\ref{sec:app_cos2}),
\begin{equation}
\nu_{T=0}^{\text{(PT)}} \approx \frac{1}{16\pi} \, \frac{E^2}{E_\text{c}^2} \, (m\tau) m^3.
\label{eq:T0_pt}
\end{equation}
The $\tau$ dependences for nonzero $T$ in Fig.~\ref{fig:switch_off} have obviously nonzero values at $\tau = 0$ and cannot be described by means of PT since the parameter $|eE|T/m$ is not small.

\begin{figure*}[t]
  \center{\includegraphics[width=0.8\linewidth]{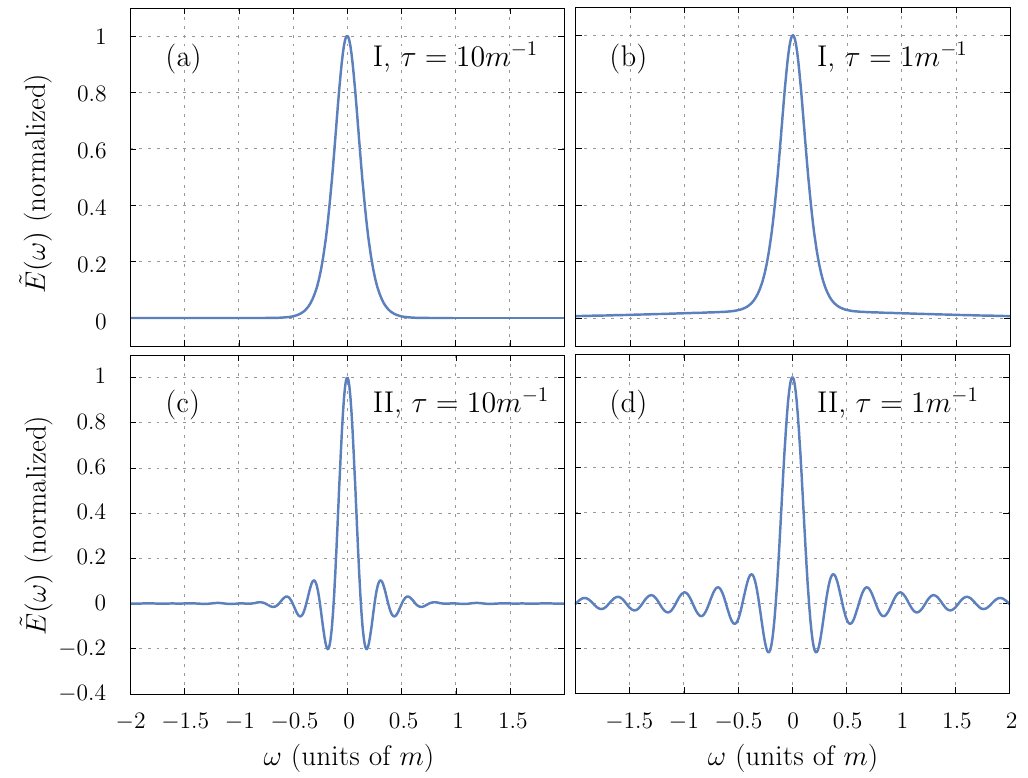}}
  \caption{Normalized Fourier transforms: (a), (b) $\tilde{E}_\text{I} (\omega)$ [see Eq.~\eqref{eq:fourier1}] for $E/\varepsilon = 4$, $T=10m^{-1}$, and two different values of $\tau$; (c), (d) $\tilde{E}_\text{II} (\omega)$ [see Eq.~\eqref{eq:fourier2}] for $T=40m^{-1}$ and two values of $\tau$.}
  \label{fig:fourier}
\end{figure*}

The LCFA does not incorporate the dynamical enhancement due to the fast switch on and off. From this it immediately follows that the field parameters must obey $|eE|^{3/2}\tau \lesssim m^2$ for the fast-switching effects to manifest themselves. Nevertheless, the inequality $|eE|^{3/2}\tau \lesssim m^2$ can be used only as a {\it necessary} condition for the pair-production enhancement. In order to identify a sufficient condition, we also employed the worldline instanton technique (see Appendix~\ref{sec:app_wli}). It turns out that the onset of the enhancement corresponds to $\gamma = m/|eE\tau| \sim 1$ similarly to the DASE condition $\gamma_\text{c} \sim 1$. This universal criterion not only reveals how the two enhancement mechanisms are similar to each other, but also allows us to assess the experimental feasibility of the fast-switch-off scenarios. These two aspects represent the main findings of our paper. Although in Figs.~\ref{fig:dase} and \ref{fig:switch_off} we discussed only examples for two specific choices of the field parameters, the criterion $\gamma \sim 1$ ($\gamma_\text{c} \sim 1$) represents now a general quantitative measure for the {\it onset} of the two enhancement mechanisms, which will be used in what follows. We also note that the condition $\gamma \gtrsim 1$ is stronger than $|eE|^{3/2}\tau \lesssim m^2$ as it should be (as in the case of the DASE, the LCFA is not able to capture the fast-switch-off enhancement).

We also point out that in Ref.~\cite{ilderton_prd_2022} it was suggested that the final particle number can be enhanced by rapidly switching off a Sauter pulse at the vicinity of its maximum. In Ref.~\cite{ilderton_prd_2022} the total number of pairs was evaluated only be means of PT. We also performed exact nonperturbative calculations using the same switching profiles as in Ref.~\cite{ilderton_prd_2022} and confirmed the main findings reported there. We found that in order to examine the main qualitative patterns of the enhancement mechanism and \sout{provide the necessary quantitative estimates} identify the enhancement onset, it suffices to explore the simple field configuration~\eqref{eq:E2} although by accurately shaping the external pulse, one may further increase the particle yield. Finally, we note that we also carried out numerical calculations in the nonsymmetric case where the switching-on part has duration $T/2$, so $\tau$ affects only the switch-off part. It was found that, at least for $T \gtrsim 10m^{-1}$, the increase in the particle yield becomes twice smaller in this case compared to the setup~\eqref{eq:E2}, which indicates that the switch-on and -off parts independently enhance pair production. The enhancement onset in the nonsymmetric case remains exactly the same.

Our findings can also be illustrated by analyzing the Fourier transforms of the corresponding external field configurations, which is the subject of the next section.

\subsection{Fourier analysis} \label{sec:fourier}

Here we will examine the Fourier transforms of the external backgrounds~\eqref{eq:E1} and \eqref{eq:E2} defined via
\begin{equation}
\tilde{E} (\omega) = \int \limits_{-\infty}^\infty E(t) \mathrm{e}^{i\omega t} dt.
\end{equation}
We find
\begin{eqnarray}
\tilde{E}_\text{I} (\omega) &=& \pi \omega \bigg [ \frac{ET^2}{\sinh \pi \omega T/2} + \frac{\varepsilon \tau^2}{\sinh \pi \omega \tau/2} \bigg ], \label{eq:fourier1}\\
\tilde{E}_\text{II} (\omega) &=& \frac{2\pi^2 E}{\omega (\pi^2 - \omega^2 \tau^2)} \, \sin  \frac{\omega (T+\tau)}{2}  \cos \frac{\omega \tau}{2}. \label{eq:fourier2}
\end{eqnarray}

\begin{figure}[t]
  \center{\includegraphics[width=0.99\linewidth]{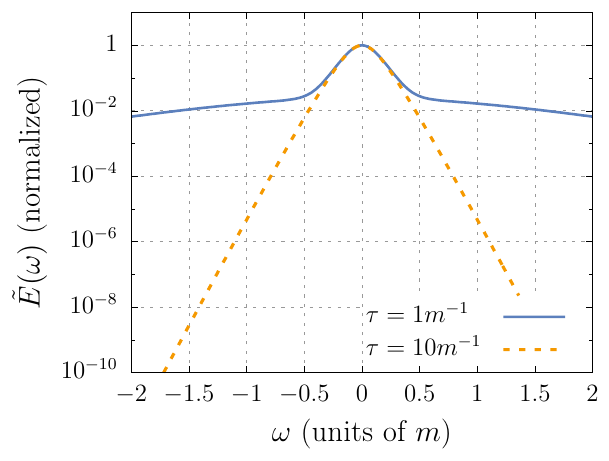}}
  \caption{Log-scale plot of the normalized Fourier transforms $\tilde{E}_\text{I} (\omega)$ presented in Fig.~\ref{fig:fourier} (a) and (b).}
  \label{fig:fourier_log}
\end{figure}

In Fig.~\ref{fig:fourier} we present the functions~\eqref{eq:fourier1} and \eqref{eq:fourier2} divided by $\tilde{E}_\text{I} (0)=2(ET + \varepsilon \tau)$ and $\tilde{E}_\text{II} (0)=E(T + \tau)$, respectively. All of the Fourier transforms are centered at $\omega = 0$ and quite narrow for relatively large $\tau$. With decreasing $\tau$, the spectra gain high-frequency contributions, which may stimulate the pair production process. For instance, in Fig.~\ref{fig:fourier}(b) the presence of a short and weak Sauter pulse alters the large-$\omega$ part of the spectrum providing it with broad wings. To evidently demonstrate this point we also present the plots in a logarithmic scale (see Fig.~\ref{fig:fourier_log}).  Although the relative contribution of the high-frequency intervals seems tiny, these harmonics lead to a significant enhancement of the pair number. In the scenario concerning a rapid switch off of the external field, the qualitative structure of the Fourier spectra is the same apart from the presence of an oscillatory pattern. For smaller values of $\tau$, one observes an increase in the high-frequency contributions. More pronounced oscillations in this case appear due to the sharp edges of the rectangular-like pulse [for $\tau = 0$ one obtains a well-known Fourier transform $\sim \sin(\omega T/2)/\omega$, which is very similar to the function displayed in Fig.~\ref{fig:fourier}(d)]. Finally, we also point out that the actual shape of the Fourier transforms may differ when considering different field profiles in the scenarios~I and~II, but the primary effect of broadening revealed for smaller values of $\tau$ will remain qualitatively the same.

As was demonstrated above, in the case of weak and rapidly varying fields, one can accurately describe the pair production process by means of PT although the main contribution in the spectrum may correspond to $\omega = 0$. This means that depending on the field parameters, the large-$\omega$ harmonics may completely govern particle production. However, within the enhancement mechanisms investigated in our study the zero-energy contribution plays also an essential role. The similarity between the Fourier transforms~\eqref{eq:fourier1} and \eqref{eq:fourier2} and between their dependence on the external field parameters indicates again that the two effects of the enhancement are indeed related. In the next section, we will discuss the experimental feasibility of the two mechanisms.

\section{Experimental prospects} \label{sec:experiment}

To find out whether the enhancement mechanisms examined in this study can be implemented experimentally, we will now evaluate the spatiotemporal scale of the rapid weak field and switch-on and -off parts, respectively. In the previous section, we provided examples of our numerical results which illustrate the two enhancement mechanisms. Although the quantitative predictions for the particle yield are rather sensitive to the field profile and its parameters chosen in specific calculations, here we will employ instead the universal conditions in terms of the Keldysh parameters $\gamma_\text{c}$ and $\gamma$, respectively.

The onset of the pair-production enhancement in the context of the DASE corresponds to $\gamma_\text{c} \equiv m/|eE\tau| = \gamma_\text{crit} \sim 1$. We note that depending on the external field profile, the critical value of the Keldysh parameter can be different. For instance, $\gamma_\text{crit} = \pi/2$~\cite{schuetzhold_prl_2008} or $\gamma_\text{crit} = \sqrt{\ln (E/\varepsilon)}$~\cite{linder_prd_2015}. Nevertheless, these values are basically 
of the order of unity, so we will assume $\gamma_\text{crit} = 1$. Let us now estimate the characteristic wavelength of the rapid weak pulse as $\lambda_\text{enh} = 2\tau$. Then by requiring $\gamma_\text{c} = 2m/|eE\lambda_\text{enh}|=1$, we find that the onset of dynamical assistance is approximately given by
\begin{equation}
\lambda_\text{enh} = 2(E_\text{c}/E)m^{-1}.
\label{eq:lambda}
\end{equation}

As was demonstrated above, by rapidly switching off the external field, one can also enhance the particle yield. The crucial point here is that the corresponding threshold is given by a similar condition $\gamma = 1$, where $\gamma = 2m/|eE\lambda_\text{enh}|$, so the corresponding length scale $\lambda_\text{enh}$ is given by exactly the same Eq.~\eqref{eq:lambda}. As was stated above, the LCFA does not have an access to the enhancement mechanisms. For completeness, we will also calculate the corresponding threshold values $\lambda_\text{LCFA} = 20 (E_\text{c}/E)^{3/2} m^{-1}$ deduced from the condition $|eE|^{3/2}\tau = 10m^2$ (see, e.g., Ref.~\cite{sevostyanov_prd_2021}).

It is known that the presence of the temporal oscillations and large volume factor $V$ effectively lower the threshold for $e^+e^-$ pair production, so that the field amplitude $E=0.1E_\text{c}$ seems already sufficient for observing the Schwinger mechanism (see, e.g., Refs.~\cite{narozhny_jetpl_2004, bulanov_jetp_2006, aleksandrov_prd_2022, tkachev_2024}). We vary $E$ from $0.001E_\text{c}$ to $0.1E_\text{c}$ and display our estimates in Table~\ref{table}. Based on the quantitative estimates obtained, we now able to assess the experimental prospects of the enhancement scenarios.

First, let us consider the enhancement mechanism due to the fast-switching effects. Since modern laser facilities generating most intense laser pulses operate in the regime of $\lambda \sim 1~\mu\text{m}$, shaping the laser field on the scale of $\lambda_\text{enh}$ appears to be \sout{completely} unrealistic. Although it is possible to generate radiation with such a short wavelength by using, e.g., free-electron laser systems, it cannot serve as a very strong background needed for the Schwinger process: this radiation should be, anyway, combined with an intense laser beam, which will have a wavelength of the level of $1~\mu\text{m}$. The attempt to {\it shape} the field on the scale much less than the laser wavelength obviously contradicts the diffraction limit. We underline that the specific features of the switching on/off effects are sensitive to the actual parameters of the pulses including the amplitude, frequency, and focusing methods, but from the experimental viewpoint this scenario is not feasible, so we turn to the analysis of the DASE.

\begin{table}[t]
\centering
\setlength{\tabcolsep}{1.2em}
\begin{tabular}{rccc}
\hline \hline
$E/E_\text{c}$ & $I$ ($\text{W}/\text{cm}^2$) & $\lambda_\text{enh}$ ($\mu$m) & $\lambda_\text{LCFA}$ ($\mu$m)   \\
\hline
  0.1  &   $2.3 \times 10^{27}$ & $8 \times 10^{-6}$ & $2 \times 10^{-4}$ \\
  0.05  &   $5.8 \times 10^{26}$ & $2 \times 10^{-5}$ & $7 \times 10^{-4}$ \\
  0.02  &   $9.2 \times 10^{25}$ & $4 \times 10^{-5}$ & $3 \times 10^{-3}$ \\
  0.01  &  $2.3 \times 10^{25}$ & $8 \times 10^{-5}$ & $8 \times 10^{-3}$ \\
  0.005  &   $5.8 \times 10^{24}$ & $2 \times 10^{-4}$ & $2 \times 10^{-2}$ \\
  0.002  &   $9.2 \times 10^{23}$ & $4 \times 10^{-4}$ & $9 \times 10^{-2}$ \\
  0.001  &   $2.3 \times 10^{23}$ & $8 \times 10^{-4}$ & $2 \times 10^{-1}$ \\
\hline \hline            
\end{tabular}
\caption{Characteristic scale (wavelength) $\lambda_\text{enh}$ of the weak pulse in the context of the DASE and the switching profile in the context of the fast-switching scenario necessary for the pair-production enhancement and the threshold value $\lambda_\text{LCFA}$ necessary for the LCFA applicability as a function of the electric field amplitude $E$ [$I = E^2/(8\pi)$ is the corresponding peak intensity]. 
\label{table}}
\end{table}

\begin{figure}[t]
  \center{\includegraphics[width=0.99\linewidth]{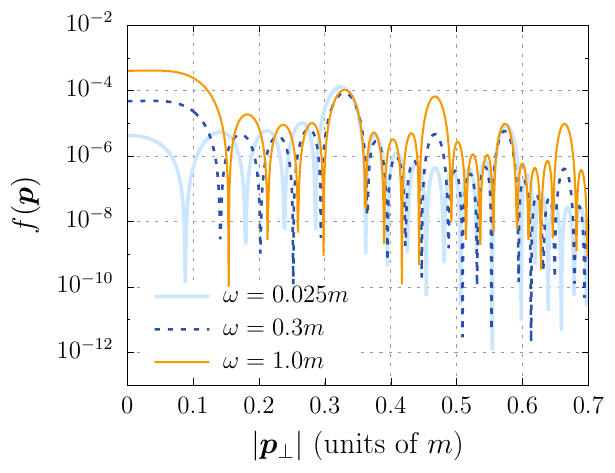}}
\caption{Momentum spectra of the particles produced within the DASE scenario~\eqref{eq:dase_osc} involving a combination of two oscillating pulses ($p_\parallel = 0$). The field parameters are $E=0.25E_\text{c}$, $\Omega = 0.025 m$, $\varepsilon = 0.01E_\text{c}$, and $N=5$, and the three curves correspond to different values of the fast-pulse frequency $\omega$.}
\label{fig:spectra}
\end{figure}

Whereas the strong field component has a wavelength of the order of $1~\mu\text{m}$, it is possible to produce relatively weak electromagnetic pulses having a much shorter wavelength. While such pulses cannot significantly change the strong-field profile, they can be superimposed on the strong component according to the idea of the DASE. Note that in this case the dynamical assistance occurs due to sufficiently large values of the weak-field frequency (small $\lambda$), while there is no need to shape the details of the field profile. This scenario should be feasible as the values of $\lambda_\text{enh}$ correspond to realistic energy scales: the data presented in Table~\ref{table} yields the photon energy of 1~keV--1~MeV. Pursuing the analysis of more realistic setups, one can now examine field configurations that differ from our simple model~\eqref{eq:E1}. To illustrate this mechanism in the case of a more realistic field, we now consider
\begin{equation}
A (t) = F(\Omega t) \bigg ( \frac{E}{\Omega} \, \sin \Omega t + \frac{\varepsilon}{\omega} \, \sin \omega t \bigg ),
\label{eq:dase_osc}
\end{equation}
where
\begin{equation}
F (\eta) = \begin{cases}
\sin^2 \big [ \frac{1}{2} (\pi N - |\eta|) \big ] &\text{if}~~\pi (N-1) \leq |\eta| < \pi N,\\
1 &\text{if}~~|t| < \pi (N-1),\\
0 &\text{otherwise}.
\end{cases}\label{eq:envelope}
\end{equation}
In Fig.~\ref{fig:spectra} we present the transverse momentum spectra for various values of the fast-pulse frequency~$\omega$ ($E=0.25E_\text{c}$, $\Omega = 0.025 m$, $\varepsilon = 0.01E_\text{c}$, $N=5$). We observe that for sufficiently large $\omega$, the particle densities are indeed significantly enhanced. According to the estimate~\eqref{eq:lambda}, the necessary value of the frequency amounts to $\omega_\text{enh} = 2 \pi/\lambda_\text{enh} \approx 0.8 m$.

The momentum distributions displayed in Fig.~\ref{fig:spectra} have a nontrivial structure. For instance, for larger $\omega$ one discovers additional intermediate peaks corresponding to the absorption of weak-field quanta~\cite{aleksandrov_prd_2018}. Moreover, the momentum patterns are also very sensitive to the spatial inhomogeneities of the external field as was thoroughly analyzed in Ref.~\cite{aleksandrov_prd_2018}. Taking into account the spatiotemporal structure of the field configuration represents here an important challenge once one strives to address more realistic setups and obtain experimentally relevant predictions. We underline here that our analysis of relatively simple field configurations demonstrates how one can identify the enhancement onset $\lambda_\text{enh}$ (or $\omega_\text{enh}$), whereas the accurate quantitative estimates of the actual enhancement factors should be then obtained by taking into account the specific structure of the corresponding experimental setup.

As can be seen from Eq.~\eqref{eq:lambda}, with increasing strong-field amplitude $E$, one has to increase also the weak-field frequency $\omega$ in order to take advantage of the dynamical assistance. It does not mean that increasing $E$ is not favorable since in stronger fields the particle yield grows with $E$ even if $\omega$ is below the DASE threshold ($\omega < \omega_\text{enh}$). The correct physical interpretation is the following: for given $E$, the individual strong-field component produces a certain amount of pairs, which can be additionally enhanced by superimposing a weak rapidly oscillating pulse. To attain an {\it efficient} dynamical enhancement, one has to make sure that the weak-field frequency $\omega$ is sufficiently high. Although the scaling $\omega_\text{enh} \sim E$ was theoretically established already in Ref.~\cite{schuetzhold_prl_2008} and confirmed in many studies including the present one, we would like to additionally illustrate it by the following simple estimates. The probability of the tunneling mechanism involving the Dirac-sea electron in the presence of a constant field $E$ can be roughly estimated as $\mathrm{exp} (-\beta L_0)$, where $L_0$ is the tunneling distance and $\beta$ is a constant of the order of $m$. The value of $L_0$ is a spatial distance where the work done by the external field amounts to $2m$, so $L_0 = 2m/|eE|$. If the electron first absorbs a high-frequency quantum from the weak field component, it significantly shortens the tunneling path: $L = (2m-\omega)/|eE| = L_0 - \omega/|eE|$. Therefore, the tunneling probability will be multiplied by the factor $\mathrm{exp} (\beta \omega/|eE|)$, which gives rise to the DASE (cf. Ref.~\cite{schuetzhold_prl_2008}). This factor contains the ratio $\omega/E$, which explains why the DASE is governed by the combined Keldysh parameter $\gamma_\text{c}$. The condition $\gamma_\text{c} \sim 1$ yields $\omega_\text{enh} \sim E$.

Finally, we point out that the parameters of the weak high-frequency pulse assisting the Schwinger mechanism might not allow one to treat it as a classical background. In this case, one should introduce quanta of the electromagnetic field (photons) in order to describe the DASE (see, e.g., Refs.~\cite{dunne_prd_2009, baier_2010, heinzl_plb_2010, karbstein_prd_2013}).

\section{Conclusion} \label{sec:conclusions}

In the present study, we investigated the dynamically assisted Schwinger effect and pair production in a strong external background being rapidly switched on and off. It was shown that both scenarios may lead to a substantial increase in the total number of pairs produced. It turns out that the two mechanisms are quite similar to each other as they both require the presence of a strong background and high-frequency harmonic. By varying the field parameters, we analyzed the onset of the pair-production enhancement and discussed the experimental prospects of the two scenarios.

Even in the case of a constant background, the adiabatic number of pairs always incorporates dynamical contribution due to a sharp switch off of the external field at the corresponding time instant. However, this initially nonphysical contribution can be used to maximize the particle yield by introducing a smooth switching function. The crucial property of this enhancement mechanism is the characteristic timescale of the switching profile. It was found that this scale is determined by the condition $m/|eE\tau| \sim 1$ having the same form as in the case of the DASE. It was shown that $\tau$ is at least three orders of magnitude smaller than the period of electromagnetic radiation having a wavelength of 1~$\mu\text{m}$. This suggests that shaping laser fields in order to take advantage of a sharp switch off is unfeasible from the experimental viewpoint.

On the other hand, one can attempt to design a setup where the pair production process occurs in a combination of a strong electric background and a rapidly oscillating field. According to our estimates, this scenario should involve high-frequency radiation with photon energy of 1~keV--1~MeV depending on the strong-field amplitude. As we were interested only in the onset of dynamical assistance, photons of higher energy may, in fact, be even more advantageous, provided the pair production process remains nonperturbative.


\begin{acknowledgments}
The study was funded by RFBR and ROSATOM, project No.~20-21-00098. I.A.A. acknowledges the support from the Foundation for the advancement of theoretical physics and mathematics ``BASIS''.
\end{acknowledgments}


\appendix

\section{Asymptotic expansion of the function~\eqref{eq:LCFA_f2}} \label{sec:app_f}

Substituting $z=\tan x$ in Eq.~\eqref{eq:LCFA_f2}, one obtains
\begin{equation}
f(\xi) = \frac{4}{\pi} \int \limits_0^{\infty} \frac{\mathrm{e}^{-\pi z^2/\xi}}{(1+z^2)^3} \, dz.
\end{equation}
This integral can be evaluated analytically in terms of the Gauss error function, but to find the small-$\xi$ expansion it suffices to integrate the Taylor series of $1/(1+z^2)^3$ multiplied by the exponential function containing $\xi$. This immediately brings us to
\begin{equation}
f(\xi) = \frac{2}{\pi} \sqrt{\xi} \Big [1 - \frac{3}{2\pi} \, \xi + \frac{9}{2\pi^2} \, \xi^2 + \mathcal{O}(\xi^3) \Big ].
\end{equation}

\begin{figure}[b]
  \center{\includegraphics[width=0.5\linewidth]{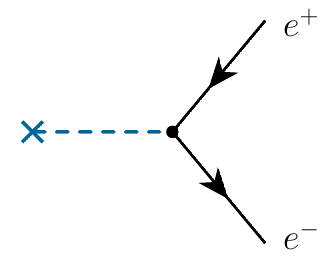}}
  \caption{Leading-order Feynman diagram describing electron-positron pair production. The line with the cross corresponds to the interaction with the external classical background. The particle number is proportional to the amplitude squared and integrated over the electron (positron) momentum.}
  \label{fig:pt}
\end{figure}

\section{Perturbation theory. Individual Sauter pulse} \label{sec:app_PT}

Let us evaluate the number of pairs to the leading (second) order in the external field amplitude. The corresponding Feynman diagram is depicted in Fig.~\ref{fig:pt}, where the cross represents the interaction with the classical background. A general expression in the case of a spatially homogeneous external field has the following form:
\begin{equation}
\nu^{\text{(PT)}} = \frac{e^2}{2(2\pi)^3}\int d\boldsymbol{p} \, \frac{m^2 + \boldsymbol{p}_\perp^2}{p_0^4} \, \Bigg | \int \limits_{-\infty}^{\infty} E(t) \mathrm{e}^{2ip_0t} \, dt \Bigg |^2,
\label{eq:app_pt_gen}
\end{equation}
where $p_0 = \sqrt{m^2 + \boldsymbol{p}^2} = \sqrt{m^2 + p_\parallel^2 + \boldsymbol{p}_\perp^2}$. To perform the calculations for a Sauter pulse $E(t) = \varepsilon/\cosh^2 (t/\tau)$, we use the Fourier transform
\begin{equation}
 \int \limits_{-\infty}^{\infty} \frac{\mathrm{e}^{i\omega t}}{\cosh^2 (t/\tau)} \, dt = \frac{\pi \omega \tau^2}{\sinh (\pi \omega \tau/2)},
\end{equation}
which can be obtained, e.g., by means of the residue theorem [it was also used in Eq.~\eqref{eq:fourier1}]. It brings us to Eq.~\eqref{eq:weak_PT}. The PT approach accurately predicts the particle yield if $\gamma_\varepsilon \equiv m/|e\varepsilon \tau| \gg 1$. Note that the integrand in Eq.~\eqref{eq:weak_PT} represents the spin-summed number density of the particles produced and coincides with Eq.~(44) from Ref.~\cite{gavrilov_prd_1996} multiplied by the spin factor $2$.

In order to derive the small-$\tau$ expression~\eqref{eq:weak_PT_lin}, we will perform the integration in terms of dimensionless $\xi = \tau p_\parallel$ and $\eta = \tau^2 |\boldsymbol{p}_\perp|^2$. One obtains
\begin{equation}
\frac{\nu^{\text{(PT)}}}{m\tau} = \frac{\varepsilon^2}{2E_\text{c}^2} \, m^3 \, \mathcal{I} (m^2 \tau^2),
\label{eq:app_I}
\end{equation}
where
\begin{equation}
\mathcal{I} (\zeta) = \int \limits_{0}^{\infty} d \xi \int \limits_{0}^{\infty} d\eta \, \frac{\eta + \zeta}{\xi^2 + \eta + \zeta} \, \frac{1}{\sinh^2 (\pi \sqrt{\xi^2 + \eta + \zeta})}.
\end{equation}
Substituting then $x = \xi^2 + \eta + \zeta$, $y = \xi^2$ and integrating over $y$ from $0$ to $x-\zeta$, we arrive at
\begin{equation}
\mathcal{I} (\zeta) = \frac{1}{3} \int \limits_{\zeta}^{\infty} \bigg ( 2 + \frac{\zeta}{x} \bigg ) \, \frac{\sqrt{x-\zeta}}{\sinh^2 (\pi \sqrt{x})} \, dx.
\end{equation}
The contribution which comes from the second term in the parentheses vanishes for $\zeta \to 0$. The value $\mathcal{I} (0)$ is nonzero due to the first term:
\begin{equation}
\mathcal{I} (0) = \frac{2}{3} \int \limits_{0}^{\infty} \frac{\sqrt{x}}{\sinh^2 (\pi \sqrt{x})} \, dx = \frac{2}{9\pi}.
\end{equation}
Together with Eq.~\eqref{eq:app_I}, it yields Eq.~\eqref{eq:weak_PT_lin}.

Finally, we point out that in the case of $1+1$ QED, there are no perpendicular components $\boldsymbol{p}_\perp$, so we have a one-dimensional integral in Eq.~\eqref{eq:weak_PT}. Instead of Eq.~\eqref{eq:weak_PT_lin}, one obtains here
\begin{equation}
\nu^{\text{(PT, 1+1)}} \approx \frac{1}{8\pi^2} \, \frac{\varepsilon^2}{E_\text{c}^2} \, (m\tau)^2 m.
\end{equation}

\section{Perturbation theory for the $\cos^2$ pulse} \label{sec:app_cos2}


The external background~\eqref{eq:E2} for $T=0$ reads $E(t) = E\cos^2 [\pi t/(2\tau)] \theta (\tau - |t|)$. First, we find
\begin{equation}
 \int \limits_{-\tau}^{\tau} \cos^2 \Big ( \frac{\pi t}{2\tau} \Big ) \, \mathrm{e}^{i\omega t} \, dt = \frac{\pi^2 \sin \omega \tau}{\omega (\pi^2 - \omega^2 \tau^2)}.
\end{equation}
This expression can be obtained from Eq.~\eqref{eq:fourier2} by setting $T=0$. Then the PT expression~\eqref{eq:app_pt_gen} can be evaluated by means of the same substitutions as those employed in Appendix~\ref{sec:app_PT}. We obtain
\begin{equation}
\frac{\nu^{\text{(PT)}}}{m\tau} = \frac{E^2}{2E_\text{c}^2} \, m^3 \, \mathcal{J} (m^2 \tau^2),
\label{eq:app_J}
\end{equation}
where
\begin{equation}
\mathcal{J} (\zeta) = \frac{\pi^2}{768} \int \limits_{\zeta}^{\infty} \frac{\sqrt{x-\zeta} ( 2x + \zeta ) \sin^2 2\sqrt{x}}{x^3 (x-\pi^2/4)^2} \, dx.
\end{equation}
This function has a finite nonzero limit as $\zeta$ tends to zero:
\begin{equation}
\mathcal{J} (0) = \frac{\pi^2}{384} \int \limits_{0}^{\infty} \frac{\sin^2 2\sqrt{x}}{x^{3/2} (x-\pi^2/4)^2} \, dx.
\end{equation}
Substituting $y = 2\sqrt{x}$, we recast this expression into
\begin{equation}
\mathcal{J} (0) = \frac{\pi^2}{6} \int \limits_{0}^{\infty} \frac{\sin^2 y}{y^2 (y^2-\pi^2)^2} \, dy = \frac{1}{8\pi}.
\label{eq:app_J0}
\end{equation}
The last integral can be computed by means of the residue theorem. Equations~\eqref{eq:app_J} and \eqref{eq:app_J0} lead to Eq.~\eqref{eq:T0_pt}.

In the case of $1+1$ QED, one obtains
\begin{equation}
\nu^{\text{(PT, 1+1)}} \approx \frac{1}{32 \pi^2} \, \frac{E^2}{E_\text{c}^2} \, (m\tau)^2 m.
\end{equation}

\section{Switch-on and -off effects via worldline instantons} \label{sec:app_wli}


Here we will employ the worldline instanton approach and demonstrate that the onset of the pair-production enhancement corresponds to $\gamma = m/|eE\tau| \gtrsim 1$.

Let us consider the external electric background in the following general form:
\begin{equation}
E(t) = E \mathfrak{e} (t/\tau),
\label{eq:app_wli_E}
\end{equation}
where $\mathfrak{e}(-z) = \mathfrak{e} (z)$, $\mathfrak{e}(0) = 1$, and $\mathfrak{e}'' (0) <0$. We also assume that the function $\mathfrak{e}(z)$ is smooth and can be considered at nonreal $z$ after a proper analytic continuation. Following Refs.~\cite{dunne_prd_2005, dunne_prd_2006}, we introduce the Euclidean gauge field
\begin{equation}
A_\text{E} (x_4) = A(ix_4) = -iE\tau f (x_4/\tau),
\end{equation} 
where
\begin{equation}
f(z) = i \mathfrak{a}(iz),\qquad \mathfrak{a} (z) = -\int \limits_0^z \mathfrak{e} (z') dz'.
\end{equation}
The function $f(z)$ is real for real $z$. If the number of $e^+e^-$ pairs is small, the particle yield is given by the imaginary part of the effective action in the external field. The latter can be represented via a worldline path integral and then computed by means of the functional stationary phase approximation (see a detailed description of this approach in Refs.~\cite{dunne_prd_2005, dunne_prd_2006}). The exponential part of the result reads:
\begin{equation}
\mathrm{exp} \bigg [ - \frac{\pi E_\text{c}}{E} \, g (\gamma) \bigg ],
\label{fig:exp_g}
\end{equation}
where
\begin{equation}
g(\gamma) = \frac{4}{\pi} \int \limits_0^{z_*(\gamma)/\gamma} \sqrt{1-\frac{1}{\gamma^2} f^2 (\gamma \xi)} \, d \xi
\label{eq:app_g}
\end{equation}
and the turning point $z_* (\gamma)$ is defined via $f(z_*) = \gamma$. Since the Taylor expansion of $f(z)$ contains only odd powers, the function $g(\gamma)$ contains only even ones. As was shown in Ref.~\cite{dunne_prd_2006}, the preexponential factor is governed by the following function including also only even powers of $\gamma$:
\begin{equation}
p(\gamma) = \frac{d}{d\gamma^2} \Big [ \gamma^2 g(\gamma) \Big ].
\end{equation}
The saddle-point approximation implies $p'(\gamma) < 0$, and the preexponential factor involves $1/\sqrt{-p'(\gamma)}$~\cite{dunne_prd_2006}. Let us use the following representation:
\begin{eqnarray}
p(\gamma) &=& \frac{2}{\pi} \int \limits_0^{z_*(\gamma)/\gamma} \frac{d\xi}{\sqrt{1-\frac{1}{\gamma^2} f^2 (\gamma \xi)}} \nonumber \\
{} &=& \frac{2}{\pi} \frac{z_*(\gamma)}{\gamma} \int \limits_0^1 \frac{d y}{\sqrt{1-\frac{1}{\gamma^2} f^2 (z_* (\gamma) y)}}.
\end{eqnarray}
The derivative reads
\begin{equation}
p'(\gamma) = -\frac{2}{\pi} \frac{1}{\gamma^3 f' (z_*)} \int \limits_0^1 \frac{f(z_*) h(z_*) - f(z_* y) h(z_* y)}{\big [1-\frac{1}{\gamma^2} f^2 (z_*  y) \big ]^{3/2}} \, dy,
\label{eq:app_pprime}
\end{equation}
where $h(z) = z f'(z) - f(z)$. In the small-$\gamma$ limit, $p(\gamma) = 1 + (1/4) \mathfrak{e}'' (0) \gamma^2 + \mathcal{O}(\gamma^4)$, so $p(\gamma)$ decreases with increasing $\gamma$. However, at a certain sufficiently large $\gamma_0$, one may encounter $p'(\gamma_0)=0$ corresponding to the breakdown of the saddle-point approximation. This can happen only if the function $h(z)$ changes the sign of its derivative, i.e. there exists $z<z_0 \equiv z_* (\gamma_0)$ where $h'(z) = 0$, i.e. $f''(z) = 0$. From Eq.~\eqref{eq:app_pprime}, it is also clear that a {\it sufficient} condition is $h(z_0) = 0$. Note that $f'(z)>0$ and $dz_*(\gamma)/d\gamma > 0$ for $z<z_0$.

\begin{figure}[b]
  \center{\includegraphics[width=0.98\linewidth]{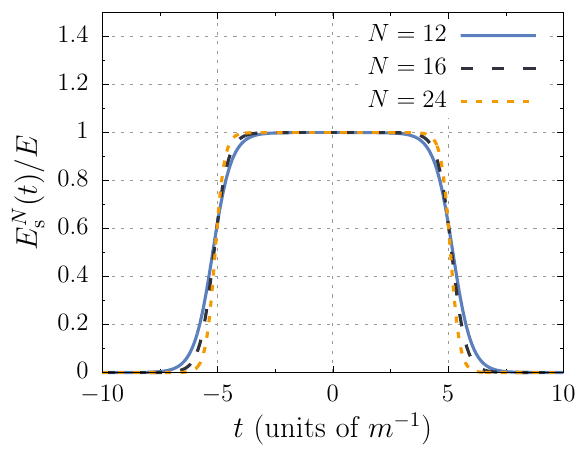}}
  \caption{Field profile~\eqref{eq:app_cosh_profile} in terms of $E_\text{s}^N (t)/E = \mathfrak{e}_\text{s}^N (t/\tau)$ as a function of $t$ for $T=10m^{-1}$ and various values of $N$ ($\tau = T/N$).}
  \label{fig:app_profile}
\end{figure}

The various field profiles considered in Ref.~\cite{dunne_prd_2006} clearly satisfy $f''(z) > 0$ within the whole range $0<z \leqslant z_* (\gamma)$ for all $\gamma >0$, so the function $p (\gamma)$ is monotonically decreasing. Namely, this is the case for the Sauter pulse $f(z) = \tan z$, Gaussian pulse $f(z) = (\sqrt{\pi}/2)\mathrm{erfi} \, z$, and oscillating cos-field $f(z) = \sinh z$. Nevertheless, a general background~\eqref{eq:app_wli_E} may correspond to a nonmonotonic function $f'(z)$. For instance, $\mathfrak{e} (z) = \mathrm{exp} [-z^2 (1 + z^2)]$ represents a bell-shaped profile quite similar to the Gaussian one but it leads to $f' (z) = \mathrm{exp} [z^2 (1 - z^2)]$ and $h(z_0)=0$ at $z_0 \approx 0.91$, so the Euclidean saddle-point approximation fails.

In order to study the switch-on and -off effects by means of the worldline instanton approach, one should first specify a smooth profile $\mathfrak{e} (z)$ with switching parts of duration $\tau$ and an extended plateau region of duration $T$. (The field configuration~\eqref{eq:E2} is very useful for QKE computations but it is not analytic even in a small vicinity of the real axis.) A natural way to generate such a profile is to utilize the antiderivative $\mathfrak{a}_\text{b} (z)$ of a {\it bell-shaped} profile $\mathfrak{e}_\text{b} (z)$ as a smooth {\it step} function and then combine two of them according to
\begin{equation}
\mathfrak{e}_\text{s} (z) = \frac{1}{2\mathfrak{a}_\text{b} (\alpha)} \, \big [ \mathfrak{a}_\text{b} (z+\alpha) - \mathfrak{a}_\text{b} (z-\alpha)\big ].
\label{eq:app_bell_step}
\end{equation}
Here we assume $E_\text{s} (t) = E \mathfrak{e}_\text{s} (t/\tau)$ and dimensionless parameter $\alpha$ determines the ratio $T/\tau$. One can straightforwardly show
\begin{eqnarray}
f_\text{s} (z) &=& - \frac{1}{\mathfrak{a}_\text{b} (\alpha)} \, \mathrm{Re} \int \limits_0^\alpha f_\text{b} (z+ix) dx, \\
f'_\text{s} (z) &=& - \frac{1}{\mathfrak{a}_\text{b} (\alpha)} \, \mathrm{Im} \, f_\text{b} (z+i\alpha).
\end{eqnarray}
It turns out that the field configurations generated this way by using exponentially decreasing bell-shaped profiles such as Sauter or Gaussian ones suffer from the problem described above. To avoid this obstacle, one can employ, for instance, $\mathfrak{e}_\text{b} (z) = 1/(1+z^2)$ but this profile does not rapidly decay, so we will proceed in a different way.

\begin{figure}[b]
  \center{\includegraphics[width=0.98\linewidth]{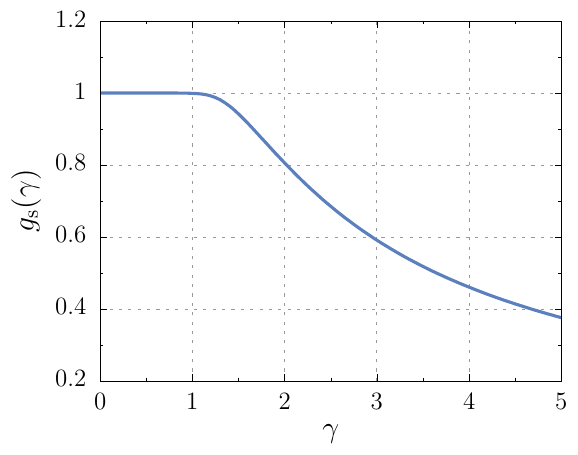}}
  \caption{Function $g_\text{s} (\gamma)$ for the smooth step profile~\eqref{eq:app_cosh_profile}.}
  \label{fig:g}
\end{figure}

Instead of using the prescription~\eqref{eq:app_bell_step}, let us construct the field profile as follows:
\begin{equation}
\mathfrak{e}^{N}_\text{s} (z) = \frac{1}{4} \sum_{n=-N}^{N} \frac{1}{\cosh^2  ( z + n/2  )},
\label{eq:app_cosh_profile}
\end{equation}
where we assume $N \gg 1$, so $\mathfrak{e}^{N}_\text{s} (0) = 1 + \mathcal{O} (\mathrm{e}^{-N}) \approx 1$. The external field $E^{N}_\text{s} (t) = E \mathfrak{e}^{N}_\text{s} (t/\tau)$ has a plateau region with duration $T = N \tau$ (see Fig.~\ref{fig:app_profile}). Although the expression~\eqref{eq:app_cosh_profile} may seem artificial from the mathematical viewpoint, it correctly mimics the physical setup under consideration and can be easily treated within the worldline instanton formalism [we explicitly verified that $p'(\gamma)$ is always negative]. One obtains
\begin{equation}
f^{N}_\text{s} (z) = \frac{1}{4}  \tan z \, \Bigg [ 1 + 2 \sum_{n=1}^{N} \frac{1 - \tanh^2 (n/2)}{1 + \tanh^2 (n/2) \tan^2 z} \Bigg ].
\end{equation}
As in the case of a single Sauter pulse, here $z_* (\gamma) \to \pi/2$ for large $\gamma$. By plugging this expression into Eq.~\eqref{eq:app_g}, we numerically evaluate the corresponding function $g^N_\text{s} (\gamma)$. It turns out that the result is almost $N$-independent, provided $N \gg 1$, so we will present the corresponding limit $g_\text{s} (\gamma)$ depending solely on $\gamma$ (see Fig.~\ref{fig:g}). We observe that the exponential part~\eqref{fig:exp_g} is substantially reduced for sufficiently large values of $\gamma$ and the corresponding threshold value amounts to $\gamma \approx 1$. Note that the shape of the function $g_\text{s} (\gamma)$ is very close to that obtained for the DASE (see Fig.~2 in Ref.~\cite{schuetzhold_prl_2008}). This is the main finding of our worldline instanton analysis.



\begin{thebibliography}{99}
%
\bibitem{sauter_1931} F.~Sauter, Z.~Phys. {\bf 69}, 742 (1931).
%
\bibitem{heisenberg_euler} W.~Heisenberg and H.~Euler, Z.~Phys. {\bf 98}, 714 (1936).
%
\bibitem{schwinger_1951} J.~Schwinger, Phys. Rev. {\bf 82}, 664 (1951).
%
\bibitem{xie_review_2017}
B.~S.~Xie, Z.~L.~Li, and S.~Tang, Matter Radiat. Extremes {\bf 2}, 225 (2017).
%
\bibitem{gonoskov_2022} A.~Gonoskov, T.~G.~Blackburn, M.~Marklund, and S.~S.~Bulanov, Rev. Mod. Phys. {\bf 94}, 045001 (2022).
%
\bibitem{fedotov_review} A.~Fedotov, A.~Ilderton, F.~Karbstein, B.~King, D.~Seipt, H.~Taya, and G.~Torgrimsson, Phys. Rep. {\bf 1010}, 1 (2023).
%
\bibitem{fedotov_prd_2011} A.~M.~Fedotov, E.~G.~Gelfer, K.~Yu.~Korolev, and S.~A.~Smolyansky, Phys. Rev. D {\bf 83}, 025011 (2011).
%
\bibitem{blaschke_prd_2013} D.~B.~Blaschke, B.~K\"ampfer, S.~M.~Schmidt, A.~D.~Panferov, A.~V.~Prozorkevich, and S.~A.~Smolyansky, Phys. Rev. D {\bf 88}, 045017 (2013).
%
\bibitem{blinne_prd_2014} A.~Blinne and H.~Gies, Phys. Rev. D {\bf 89}, 085001 (2014).
%
\bibitem{zahn_2015} J. Zahn, J. Phys. A {\bf 48}, 475402 (2015).
%
\bibitem{otto_jpp_2016} A.~Otto, T.~Nousch, D.~Seipt, B.~K\"ampfer, D.~Blaschke, A.~D.~Panferov, S.~A.~Smolyansky, and A.~I.~Titov, J. Plasma
Phys. {\bf 82}, 655820301 (2016).
%
\bibitem{ilderton_prd_2022} A.~Ilderton, Phys. Rev. D {\bf 105}, 016021 (2022).
%
\bibitem{schuetzhold_prl_2008} R.~Sch\"utzhold, H.~Gies, and G.~Dunne, Phys. Rev. Lett. {\bf 101}, 130404 (2008).
%
\bibitem{orthaber_2011} M.~Orthaber, F.~Hebenstreit, and R.~Alkofer, Phys. Lett. B
{\bf 698}, 80 (2011).
%
\bibitem{fey} C.~Fey and R.~Sch\"utzhold, Phys. Rev. D {\bf 85}, 025004 (2012).
%
\bibitem{kohlfuerst_prd_2013} C.~Kohlf\"urst, M.~Mitter, G.~von Winckel, F.~Hebenstreit, and R.~Alkofer, Phys. Rev. D {\bf 88}, 045028 (2013).
%
\bibitem{akal_prd_2014} I.~Akal, S.~Villalba-Ch\'avez, and C.~M\"uller, Phys. Rev. D {\bf 90}, 113004 (2014).
%
\bibitem{hebenstreit_plb_2014} F.~Hebenstreit and F. Fillion-Gourdeau, Phys. Lett. B {\bf 739}, 189 (2014).
%
\bibitem{linder_prd_2015} M.~F.~Linder, C.~Schneider, J.~Sicking, N.~Szpak, and R.~Sch\"utzhold, Phys. Rev. D {\bf 92}, 085009 (2015).
%
\bibitem{otto_plb_2015} A.~Otto, D.~Seipt, D.~Blaschke, B.~K\"ampfer, and S.~A.~Smolyansky, Phys. Lett. B {\bf 740}, 335 (2015).
%
\bibitem{panferov_epjd_2016} A.~D.~Panferov, S.~A.~Smolyansky, A.~Otto, B.~K.~K\"ampfer, D.~B.~Blaschke, and L.~Juchnowski, Eur. Phys. J. D {\bf 70}, 56 (2016).
%
\bibitem{aleksandrov_prd_2018} I.~A.~Aleksandrov, G.~Plunien, and V.~M.~Shabaev, Phys. Rev. D {\bf 97}, 116001 (2018).
%
\bibitem{otto_epja_2018} A.~Otto, H.~Oppitz, and B.~K\"ampfer, Eur. Phys. J. A {\bf 54}, 23 (2018).
%
\bibitem{sitiwaldi_plb_2018} I.~Sitiwaldi and B.-S.~Xie, Phys. Lett. B {\bf 777}, 406 (2018).
%
\bibitem{torgrimsson_prd_2018} G.~Torgrimsson, C.~Schneider, and R.~Sch\"utzhold, Phys. Rev. D {\bf 97}, 096004 (2018).
%
\bibitem{huang_prd_2019} X.-G.~Huang and H.~Taya, Phys. Rev. D {\bf 100}, 016013 (2019).
%
\bibitem{chavez_prd_2019} S.~Villalba-Ch\'avez and C.~M\"uller, Phys. Rev. D {\bf 100}, 116018 (2019).
%
\bibitem{olugh_plb_2020} O.~Olugh, Z.~L.~Li, B.-S.~Xie, Phys. Lett. B {\bf 802}, 135259 (2020).
%
\bibitem{taya_prr_2020} H.~Taya, Phys. Rev. Res. {\bf 2}, 023257 (2020).
%
\bibitem{li_prd_2021} L.-J.~Li, M.~Mohamedsedik, and B.-S.~Xie, Phys. Rev. D {\bf 104}, 036015 (2021).
%
\bibitem{aleksandrov_qke_2020} I.~A.~Aleksandrov, V.~V.~Dmitriev, D.~G.~Sevostyanov, and S.~A.~Smolyansky, Eur. Phys. J. Spec. Top. {\bf 229}, 3469 (2020).
%
\bibitem{sevostyanov_prd_2021} D.~G.~Sevostyanov, I.~A.~Aleksandrov, G.~Plunien, and V.~M.~Shabaev, Phys. Rev. D {\bf 104}, 076014 (2021).
%
\bibitem{aleksandrov_prd_2016} I.~A.~Aleksandrov, G.~Plunien, and V.~M.~Shabaev, Phys. Rev. D {\bf 94}, 065024 (2016).
%
\bibitem{aleksandrov_prd_2017_2} I.~A.~Aleksandrov, G.~Plunien, and V.~M.~Shabaev, Phys. Rev. D {\bf 96}, 076006 (2017).
%
\bibitem{aleksandrov_kohlfuerst} I.~A.~Aleksandrov and C.~Kohlf\"urst, Phys. Rev. D {\bf 101}, 096009 (2020).
%
\bibitem{dunne_prd_2005} G.~V.~Dunne and C.~Schubert, Phys. Rev. D {\bf 72}, 105004 (2005).
%
\bibitem{dunne_prd_2006} G.~V.~Dunne, Q.~H.~Wang, H.~Gies, and C.~Schubert, Phys. Rev. D {\bf 73}, 065028 (2006).
%
\bibitem{bunkin_tugov} F.~V.~Bunkin and I.~I.~Tugov, Dokl. Akad. Nauk SSSR {\bf 187}, 541 (1969) [Sov. Phys. Dokl. {\bf 14}, 678 (1970)].
%
\bibitem{narozhny_pla_2004} N.~B.~Narozhny, S.~S.~Bulanov, V.~D.~Mur, and V.~S.~Popov, Phys. Lett. A {\bf 330}, 1 (2004).
%
\bibitem{hebenstreit_prdr_2008} F.~Hebenstreit, R.~Alkofer, and H.~Gies, Phys. Rev. D {\bf 78}, 061701(R) (2008).
%
\bibitem{bulanov_prl_2010} S.~S.~Bulanov, V.~D.~Mur, N.~B.~Narozhny, J.~Nees, and V.~S.~Popov, Phys. Rev. Lett. {\bf 104}, 220404 (2010).
%
\bibitem{gavrilov_prd_2017} S.~P.~Gavrilov and D.~M.~Gitman, Phys. Rev. D {\bf 95}, 076013 (2017).
%
\bibitem{aleksandrov_prd_2019_1} I.~A.~Aleksandrov, G.~Plunien, and V.~M.~Shabaev, Phys. Rev. D {\bf 99}, 016020 (2019).
%
\bibitem{aleksandrov_symmetry_2022} I.~A.~Aleksandrov, D.~G.~Sevostyanov, and V.~M.~Shabaev, Symmetry {\bf 14}, 2444 (2022).
%
\bibitem{GMM} A.~A.~Grib, S.~G.~Mamaev, and V.~M.~Mostepanenko, {\it Vacuum Quantum Effects in Strong External Fields} (Friedmann Laboratory Publishing, St. Petersburg, 1994).
%
\bibitem{schmidt_1998} S.~Schmidt, D.~Blaschke, G.~R\"opke, S.~A.~Smolyansky, A.~V.~Prozorkevich, and V.~D.~Toneev, Int. J. Mod. Phys. E {\bf 07}, 709 (1998).
%
\bibitem{kluger_1998} Y.~Kluger, E.~Mottola, and J.~M.~Eisenberg, Phys. Rev. D {\bf 58}, 125015 (1998).
%
\bibitem{pervushin_2006} V.~N.~Pervushin and V.~V.~Skokov, Acta Phys. Pol. B {\bf 37}, 2587 (2006).
%
\bibitem{filatov_2006} A.~V.~Filatov, A.~V.~Prozorkevich, and S.~A.~Smolyansky, Proc. SPIE Int. Soc. Opt. Eng. {\bf 6165}, 616509 (2006).
%
\bibitem{aleksandrov_kudlis_klochai_2024} I.~A.~Aleksandrov, A.~Kudlis, and A.~I.~Klochai, Phys. Rev. Res. {\bf 6}, 043009 (2024).
%
\bibitem{nikishov_constant} A.~I.~Nikishov, Zh. Eksp. Teor. Fiz. {\bf 57}, 1210 (1969) [Sov. Phys. JETP {\bf 30}, 660 (1970)].
%
\bibitem{lcfa_comment} There is still a subtle discrepancy for large $\tau$ since the LCFA is not completely justified even if the weak pulse is absent: for the strong-pulse parameters chosen here and $\varepsilon=0$, the LCFA yields $4.0 \times 10^{-10} m^3$, while the exact number of pairs produced by the strong pulse amounts to $5.4 \times 10^{-10} m^3$ (see the horizontal line in Fig.~\ref{fig:dase}). For an individual Sauter pulse, the LCFA accurately predicts the particle yield if the parameter $|eE|^{3/2} T$ is much larger than $m^2$~\cite{sevostyanov_prd_2021}, but in our case $|eE|^{3/2} T \approx 1.8 m^2$.
%
\bibitem{kohlfuerst_arxiv_2022} C.~Kohlf\"urst, N.~Ahmadiniaz, J.~Oertel, and R.~Sch\"utzhold, Phys. Rev. Lett. {\bf 129}, 241801 (2022).
%
\bibitem{narozhny_jetpl_2004} N.~B.~Narozhny, S.~S.~Bulanov, V.~D.~Mur, and V.~S.~Popov, JETP Lett. {\bf 80}, 382 (2004).
%
\bibitem{bulanov_jetp_2006} S.~S.~Bulanov, N.~B.~Narozhny, V.~D.~Mur, and V.~S.~Popov, JETP {\bf 102}, 9 (2006).
%
\bibitem{aleksandrov_prd_2022} I.~A.~Aleksandrov, A.~Di~Piazza, G.~Plunien, and V.~M.~Shabaev, Phys. Rev. D {\bf 105}, 116005 (2022).
%
\bibitem{tkachev_2024} A.~G.~Tkachev, I.~A.~Aleksandrov, and V.~M.~Shabaev, arXiv:2408.04084.
%
\bibitem{dunne_prd_2009} G.~V.~Dunne, H.~Gies, and R.~Sch\"utzhold, Phys. Rev. D {\bf 80}, 111301(R) (2009).
%
\bibitem{baier_2010} V.~N.~Baier and V.~M.~Katkov, Phys. Lett. A {\bf 374}, 2201 (2010).
%
\bibitem{heinzl_plb_2010} T.~Heinzl, A.~Ilderton, and M.~Marklund, Phys. Lett. B
{\bf 692}, 250 (2010).
%
\bibitem{karbstein_prd_2013} F.~Karbstein, Phys. Rev. D {\bf 88}, 085033 (2013).
%
\bibitem{gavrilov_prd_1996} S.~P.~Gavrilov and D.~M.~Gitman, Phys. Rev. D {\bf 53}, 7162 (1996).
%
\end{thebibliography}
\end{document}